\documentclass[doublecol]{epl2}
\usepackage[utf8]{inputenc}
\usepackage{graphicx,color}
\usepackage{times}
\usepackage{epsf}
\usepackage[colorlinks={true}]{hyperref}
\hypersetup{citecolor={blue}, filecolor={blue}, linkcolor={blue}, urlcolor={blue}}
\title{A photonic Carnot engine powered by a quantum spin-star network}
\shorttitle{Spin interaction based work extraction} 
\author{Deniz T\"{u}rkpen\c{c}e\inst{1}, Ferdi Altintas\inst{2,3}, Mauro Paternostro\inst{4},
\and  \"{O}zg\"{u}r~E.~M\"{u}stecapl{\i}o\u{g}lu\inst{2}}
\shortauthor{D.~T\"{u}rkpen\c{c}e, F.~Altintas, M.~ Paternostro,  and \"{O}~ E.~M\"{u}stecapl{\i}o\u{g}lu}
\institute{ \inst{1} Bahcesehir College, Haskoy, 170. Sk. 55800, Canik-Samsun, Turkey \\
  \inst{2} Department of Physics, Ko\c{c} University, Sar{\i}yer, \.Istanbul, 34450, Turkey\\
  \inst{3} Department of Physics, Abant Izzet Baysal University, Bolu, 14280, Turkey\\
  \inst{4} Centre for Theoretical Atomic, Molecular and Optical Physics, School of Mathematics and Physics, 
               Queen's University, Belfast BT7 1NN, United Kingdom
}
\abstract{ We propose a spin-star network, where a central spin-$1/2$ is coupled with XXZ interaction to 
$N$ outer spin-$1/2$ particles, as a quantum fuel. 
If the network is in thermal equilibrium with a cold
bath, the central spin can have an effective temperature larger than the bath one and 
scaling nonlinearly with $N$. The nonlinearity can be tuned to $N^2, N^3$ or $N^4$
with the anisotropy parameter of the coupling. Using  a stream of central-spin particles to pump
a micromaser cavity, we calculate the dynamics of the cavity field using a coarse-grained master
equation. Our study reveals
that the central-spin beam effectively acts as a hot reservoir to the cavity field and brings the field to a thermal steady-state
whose temperature benefits from the same nonlinear enhancement with $N$, and results in a highly efficient photonic Carnot engine. The validity of our conclusions is tested against the presence of atomic and cavity damping using a microscopic master equation method for typical microwave cavity-QED parameters. An alternative equivalent scheme where the spin-$1/2$ is coupled to a macroscopic spin-$(N/2)$ particle is also discussed. 
}
\begin{document}
\maketitle
\section{Introduction}
Quantum heat engines (QHEs) are energy harvesting machines with quantum working substances, which allow for exploring the relations
among energy, coherence, and correlations in quantum thermodynamics~\cite{geva_quantummechanical_1992, feldmann_performance_2000, dillenschneider_energetics_2009, li_quantum_2014,huang_effects_2012, rosnagel_nanoscale_2014,
altintas_general_2015, turkpence_quantum_2016, quan_quantum-classical_2006, quan_quantum_2007, scully_extracting_2003, azimi_quantum_2014,altintas_quantum_2014, altintas_rabi_2015, cakmak_lipkin-meshkov-glick_2016}. 
A particularly intriguing class of QHEs can harvest quantum coherent resources to produce useful work in a single thermal 
environment and without violating the second law~\cite{scully_extracting_2003}. 
A major obstacle against the realization of such QHEs in real systems is quantum decoherence, 
under which QHEs become classical~\cite{quan_quantum-classical_2006}. Quantum decoherence can be beaten
by critical scaling of quantum coherence in multi-level atoms~\cite{turkpence_quantum_2016} or 
by using superradiant atomic clusters~\cite{hardal_superradiant_2015}. Such schemes 
however involve sophisticated coherence-injection schemes with high energy cost. Moreover, the coherence needs to be small 
for quasi-thermal 
equilibrium operation, which makes the efficiency of the engine low. {While the low-coherence condition can be 
relaxed using only those
quantum coherences that can be exchanged to heat exclusively~\cite{dag_multiatom_2016},} finding simple and 
efficient preparation schemes of sizeable initial coherences still remains a challenge against an efficient QHE.
 
{A photonic Carnot engine (PCE) uses a photon gas in an optical cavity as a working substance. The cavity is pumped by an 
atomic beam in the isothermal expansion stage~\cite{scully_extracting_2003}. Here, we investigate a generalization 
of the idea of 
forming the atomic beam by sending only one atom of a pair of interacting two-level atoms 
in thermal equilibrium through the cavity~\cite{dillenschneider_energetics_2009}.} 
The single-atom fuel remains in a thermal state so that the weak-coherence constraint is relaxed. 
The coherence is internally generated by interatomic interactions and characterized by thermal entanglement or quantum discord,
which are closely related to the energetics of the system~\cite{dillenschneider_energetics_2009,anders_thermal_2008}.  
Thermal entanglement was originally proposed as a form of pairwise quantum correlation between spins in natural anti-ferromagnets
at thermal equilibrium, which can be enhanced by using a magnetic field and putting the system in contact 
with a thermal reservoir~\cite{arnesen_natural_2001}. Quantum
discord quantifies the degree of genuinely quantum correlations in the state of a system and results from the subtraction of the maximum degree of classical correlations from the
 total correlations (quantum and classical) quantified by the mutual information~\cite{ollivier_quantum_2001}.
Magnetic susceptibility
and heat capacity experiments have been used to verify the presence of entanglement~\cite{chakraborty_signature_2014} and discord~\cite{singh_experimental_2015} in thermal states. The possibility of finding such quantum correlations in nature and
their {robustness~\cite{vidal_robustness_1999,markham_survival_2008,brukner_crucial_2006,anders_detecting_2006,soares-pinto_entanglement_2009}} make them appealing resources for QHEs fuelled by thermal states. 

\begin{figure*}[!ht]
            \includegraphics[width=18cm]{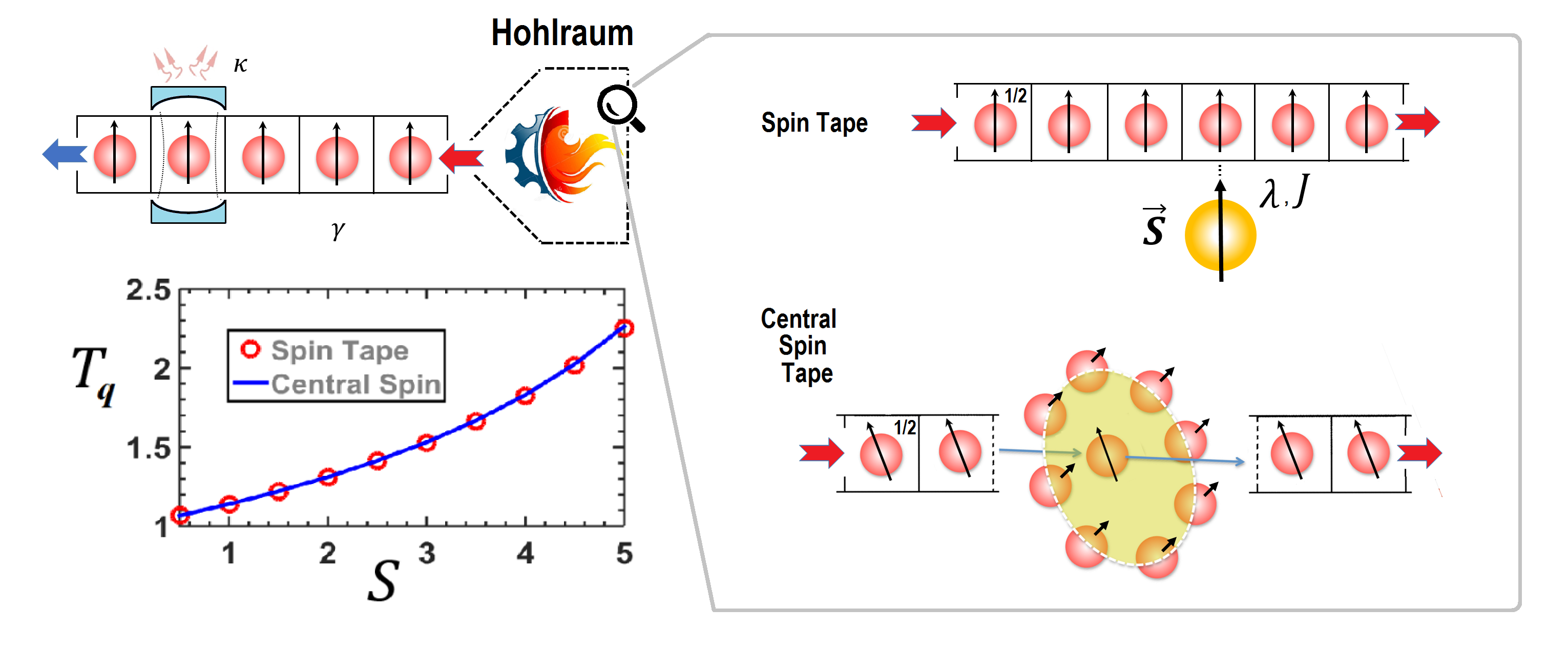}
                \caption{ Micromaser cavity as a photonic Carnot engine (PCE) powered by a tape of spin-$1/2$ particles (top left). 
                Spin-tape particles are initialized with either of two schemes (top and bottom of the right panel), which lead to the
                same local temperature $T_q$ of the spin-tape particles (bottom left). The magnitude $S$ of spin-$\vec{S}$ and
                the number of the outer spins in the central spin network are related to each other via $N=2S$. 
                The parameters used for the plot are the Bohr frequency
                of the spins $\omega=6.0$, spin-spin interaction coefficient $J=0.8$ and anisotropy ratio $\lambda=0.75$. 
                All the parameters are dimensionless. The Bohr frequency $\omega$, $T_q$, and spin-spin interaction coefficient 
                $J$ are scaled with $T_b$ (we have taken $\hbar=k_B=1$). 
                The cavity loss $\kappa$ and the atomic decay $\gamma$ are ignored
                in the plot.}
   \label{fig:fig1}
\end{figure*}

We consider a spin-star network where a centre spin is isotropically and homogeneously coupled to a 
collection of outer ones (cf. Fig.~\ref{fig:fig1}). The central spin serves as an ancillary thermal fuel for a PCE. {As such system is effectively in a thermal state, it can exchange large amount of quantum coherence, scaling as functions of the number of outer spins, exclusively as heat with the cavity field of the engine.
Instead of sophisticated optical schemes for the injection of coherence, which are typically energetically very costly, natural interactions among the spins in the 
spin-star network are used to generate quantum coherence through thermal entanglement.} 
In this way, we expect to retrieve previously identified scaling-induced quantum advantages in a relatively simple scheme for 
quantum-fuel engineering. In addition, our scheme paves the way for high temperature QHEs that can harvest efficiently sizeable ``natural''
quantum resources in a single thermal environment.

A spin-star network is a special case of Gaudin's central spin~\cite{breuer_non-markovian_2004} or Coleman-Hepp models~\cite{hepp_quantum_1972}. Such models can be used for systematic studies of quantum to classical transition~\cite{zurek_decoherence_2003,cucchietti_decoherence_2005}
and quantum decoherence~\cite{hiyama_generalized_1993}. They can be used to describe nuclear spin baths in
Nitrogen-Vacancy center defects in diamond~\cite{hanson_coherent_2008}, semiconductor 
quantum dots~\cite{imamoglu_quantum_1999,bluhm_dephasing_2011}, 
and nuclear magnetic resonance systems~\cite{zhang_detection_2008}. spin-star network is a simplified model as the outer
spin bath is assumed non-interacting. Another application area of the spin-star networks is quantum clonning and entanglement
distribution for quantum communication~\cite{de_chiara_quantum_2004,chen_optimal_2006,hutton_comparison_2002}.  
In contrast to typical motivation of investigating environmental
influence on the central spin coherence, our purpose is to examine effects of outer spins on the thermal character of the central one.

A closed spin-star network is a strongly non-Markovian system and outer spins in thermal state cannot thermalize the central 
one~\cite{guo_dissipative_2014, breuer_non-markovian_2004 }.
We consider an open network in thermal equilibrium with a hohlarum at temperature $T_b$. In this case, the outer spins
are not in a thermal state while the central spin can be in an effective thermal state, at a different effective temperature than $T_b$.
For two spins, the temperature difference is attributed to quantum entanglement at low temperatures and to quantum 
discord at higher temperatures~\cite{dillenschneider_energetics_2009}. Zero temperature spin-star network exhibits entanglement
which is scaled with the 
number of outer spins $N$~\cite{hutton_mediated_2004}. Thermal entanglement in spin-star network with few outer spins has been
shown~\cite{wan-li_tunable_2009,militello_genuine_2011}. Motivated by these results~\cite{dillenschneider_energetics_2009,hutton_mediated_2004,wan-li_tunable_2009,militello_genuine_2011}, we predict that
scaling of the correlations with the number of the outer spins can survive at higher temperatures and can be translated to 
the temperature of the central spin. 
Specifically, we consider a generalized spin 
star model with Heisenberg XXZ interaction to test our predictions. This particular model is popular in quantum communication~\cite{chen_optimal_2006} and existence of the tunable thermal entanglement is known~\cite{wan-li_tunable_2009}.
Our analysis reveals that temperature of the central spin can indeed exhibit non-trivial scaling and enhancement with the number of
outer spins and it can be tuned by the system parameters, in particular with the anisotropic coupling coefficient. 
We propose to use enhancement of the
temperature of the central spin relative to the environment to power up a PCE in the same thermal environment, generalizing the idea
for the case of two qubits~\cite{dillenschneider_energetics_2009}. 

In addition to a spin-star network, we consider an alternative set-up where a single spin-$1/2$ is coupled to a spin-$S$,
which can be compared to a similar set up of a spin-tape coupled to a quantum-dot spin valve~\cite{strasberg_second_2014} 
(cf. Fig.~\ref{fig:fig1}). In this case, a spin-tape consisting of spin-$1/2$ particles sequentially interact with 
a spin-$S$ ($S= 1/2, 1, 3/2,..$) particle for a time sufficient to thermalize the total system to $T_b$. The spin-$S$ is 
locally in a non-equilibrium state while the spin-$1/2$
is in a thermal state with local temperature that can be different than $T_b$~\cite{altintas_general_2015}. 
We find that this case is equivalent to spin-star scheme from the perspective of local temperature of tape particles,
which are considered to be quantum fuel for the PCE, and may be easier to implement. Using a single spin-$S$, 
instead of a spin-star network, allows for more compact realizations, for example by molecular nanomagnets~\cite{timco_engineering_2009}, 
and access to wider range of spin values. 
Moreover it alleviates the requirement of homogeneous 
coupling between the central spin and the environment spins
Existence of thermal entanglement in this
model has already been reported~\cite{li_thermal_2012}. 

After the interaction with the ancillary spin enviornment inside the heat reservoir, central spins are sent through an 
optical resonator, with random arrival times  (cf. Fig.~\ref{fig:fig1}). The working dynamics of the PCEs powered by the schemes 
depicted in Fig.~\ref{fig:fig1} can be described by master equations, similar to those of 
stochastic Turing machines~\cite{strasberg_thermodynamics_2015}.
We perform numerical simulations of proposed PCE with 
decoherence channels, including spontaneous atom decay $\gamma$ and cavity loss $\kappa$ using typical parameters of microwave
cavity QED systems~\cite{blais_cavity_2004,pielawa_engineering_2010}. Another promising system for realization could be
transmon qubits and transmission line resonators~\cite{heras_digital_2014,makhlin_josephson-junction_1999}. 
Quantum dot systems~\cite{wiele_quantum-dot_1999,strasberg_second_2014} or 
coupled microcavities~\cite{wan-li_tunable_2009,hartmann_effective_2007} can be
considered as well for implementations. Finite size scaling with the number of spins and the effect of thermal entanglement have been discussed in a working substance of multiferroic
chain of a quantum Otto engine and no significant dependence, except an abrupt change at three working spins, has been found in the efficiency of the engine~\cite{azimi_quantum_2014}. Our investigations here are about the finite size scaling 
in the quantum reservoirs of a PCE.  
We numerically verify our theoretical predictions that spin-star powered PCE can reach high efficiency values even for 
a single thermal environment under realistic conditions.
\section{Initialization}
We consider two different spin environments to initialize spin-$1/2$ particles before injecting them into a 
micromaser cavity~\cite{filipowicz_theory_1986}, as a fuel for a PCE, 
as shown in Fig.~\ref{fig:fig1}. The top of the right panel of Fig.~\ref{fig:fig1} describes the case, where 
spin-$1/2$ particles in a spin-tape interact sequentially with a spin-$S$ particle at rest. 
In the other case, depicted in the bottom of the right panel of Fig.~\ref{fig:fig1}, 
spin-tape particles interact sequentially with $N=2S$ number of spin-$1/2$ particles, which are positioned as outer spins of a 
spin-star network. The interactions in both cases happen inside a {\it hohlarum} at a temperature $T_b$. After sufficient time for thermalization of the total spin system to $T_b$, the tape spins are carried out of the hohlarum into the micromaser cavity, 
as shown in the top left of Fig.~\ref{fig:fig1}.

For both schemes the interaction between the tape spins and the stationary spins is described by the Heisenberg $XXZ$
model, which is usually considered for quantum communication applications of spin-star network~\cite{chen_optimal_2006}  and its thermal entanglement properties are established~\cite{wan-li_tunable_2009}. The Hamiltonian of the spin-star system is 
\begin{eqnarray}\label{H2}
H=\frac{\omega}{2}\sum_{i=0}^N \sigma_{iz}+\frac{J}{4}\sum_{i=1}^N (\sigma_{0x}\sigma_{ix}+\sigma_{0y}\sigma_{iy}+\lambda \sigma_{0z}\sigma_{iz}),
\end{eqnarray}
where the operators of the outer spins are denoted by $\sigma_{i\alpha}$ with $i=1..N$. In the case of single spin-$S$ environment, the model Hamiltonian is given by 
(we have taken $\hbar = 1$)~\cite{altintas_general_2015}
\begin{eqnarray}\label{H1}
H=\frac{\omega}{2} \sigma_{0z}+\omega S_{1z}+\frac{J}{2}\left(\sigma_{0x}S_{1x}+\sigma_{0y}S_{1y}+\lambda \sigma_{0z}S_{1z}\right).
\end{eqnarray}
Here, $\omega$ is the Bohr frequency, $J$ is the anti-ferromagnetic coupling strength, $\lambda$ is the anisotropy parameter in the exchange coupling in the $z$-direction. The operators of the tape spins and the stationary spin are denoted by $\sigma_{0\alpha}$ and $S_{1\alpha}$ along directions $\alpha=x,y,z$, respectively. Identifying
$S_{1\alpha}$ as the collective spin operators  $S_{1\alpha}=(1/2)\sum_i\sigma_{i\alpha}$, the two models can be considered as
equivalent to each other. The local temperatures of the tape spins for both models are identical at $S=N/2$; and nonlinearly vary with 
$S$ as shown in the bottom left of Fig.~\ref{fig:fig1}. 

The density matrix of the total system for both cases, which is in thermal equilibrium with a heat bath at temperature $T_b$, 
can be given by the Gibbs canonical distribution,
$\rho=\exp{(-\beta_b H)}/Z$,
where $Z=\mathrm{Tr}[\exp{(-\beta_b H)}]$ is the partition function and $\beta_b=1/T_b$ is the inverse bath temperature 
(we have taken $k_B=1$). The reduced density matrix $\rho_0$ of a tape spin is obtained by tracing out the other spin 
degrees of freedom 
and for both cases it is given by 
\begin{eqnarray}\label{rqdm}
\rho_{0}=p_e\left|e\right\rangle\left\langle e\right|+(1-p_e)\left|g\right\rangle\left\langle g\right|,
\end{eqnarray}
where $\left|e\right\rangle$ $\left(\left|g\right\rangle\right)$ is the excited (ground) state of the spin-$1/2$ and $p_e$ $(p_g=1-p_e)$ 
is the corresponding occupation probability. For $p_e<1/2$, a finite positive effective temperature for the tape spin can
be defined as $T_{q}=-\omega/\ln(p_e/p_g)$ with $p_e=\exp(-\omega/2T_q)/Z_0$ and $Z_0=2\cosh{(\omega/2T_q)}$. For the non-interacting case ($J=0$), we always have $T_q=T_b$; 
while $T_q$ can be different than $T_b$ in the  presence of interactions. For an interacting case, the enhancement of the effective
temperature exhibits a non-linear
scaling with the number of outer spins in a spin-star network (or equivalently with the spin-$S$) as shown in the bottom left plot
in Fig.~\ref{fig:fig1}, which can be exploited to power up a PCE.
\section{Photonic Carnot engine}
\begin{table*}[tbp]\centering
{
\begin{tabular}{| c | c | c | c | c | c | c |}
\hline
 $\mathcal{\lambda}$  & $a$ &  $b$  & $c$ & $d$ & $e$ &  $R^2$ \\ \hline
 $0.0$  & 1.0030 & 0.0236 & 0.0 & 0.0 & 0.0 & 0.9995 \\ \hline
 $0.25$  & 0.9856  & 0.0708 & 0.0 & 0.0 &  0.0 & 0.9975  \\ \hline
 $0.5$  & 1.0215  & 0.0637 & 0.0138 & 0.0 &  0.0 &  0.9995  \\ \hline
 $0.75$ & 0.9552 & 0.2212 & -0.0394 & 0.0096 & 0.0 &  0.9997  \\ \hline
 $0.9$ & 0.8189 & 0.5149 & -0.1719 & 0.0306 & 0.0 &   0.9982  \\ \hline
 $1.0$ & 1.3518 & -0.7877 & 0.7572 & -0.2138 & 0.0222 &   0.9989  \\ \hline
 
\end{tabular}
\caption{The fitting parameters to the data in Fig.~2. A featureless non-linear polynomial $y=a+bx+cx^2+dx^3+ex^4$ is used as a fitting curve to show the non-linear grow of the cavity temperature as a function of spin-$S$ value. The fitting for each $\lambda$ is intended to be performed with less parameters and how better the non-linear model fits the data is determined with $R^2$ values.}
\label{table1}
}
\end{table*}

A PCE~\cite{scully_extracting_2003} consists of a single mode micromaser cavity field as a working substance undergoing a 
quantum Carnot cycle consisting of two quantum adiabatic and two isothermal processes~\cite{quan_quantum_2007}. The
pump atoms injected into the cavity serve as an effective hot reservoir, while the
 environment of the system provides a cold bath. 
 Radiation pressure of the cavity field performs work on one of the cavity mirrors playing the role
 of a mechanical piston. 
 
 In contrast to the original proposal of PCE in which the pump atoms are in quasi-thermal states with weak coherence 
~\cite{scully_extracting_2003}, we consider pump atoms as the spin-$1/2$ particles 
coming out of an interacting spin ensemble at thermal equilibrium in a hohlarum at $T_b$, as shown in Fig.~\ref{fig:fig1}. 
The cavity field interacts with
tape spins in effective thermal states without any coherence. Effective temperature $T_q$ of the atoms however depends on the
number of spins in the ensemble as well as interaction parameters; and it can be different than
the $T_b$. This reflects the fact that the total thermal state can have quantum correlations and subsystems are 
in general non-equilibrium states~\cite{dillenschneider_energetics_2009}. 

During the isothermal expansion process of the Carnot cycle, in which the pump spins are sent through the cavity, the
field remains in a thermal steady state at temperature $T_q$, due to the variation of the cavity field frequency. 
For $p_e<1/2$, the field in a micromaser is exactly thermal~\cite{scully_quantum_1967}. Assuming the cold bath is at the same 
temperature $T_b$ with the hohlarum, the thermal efficiency of the PCE can be given as the generalized Carnot efficiency
$\eta=1-T_b/T_q$.
\begin{figure}[!ht]
            \includegraphics[width=8.6cm]{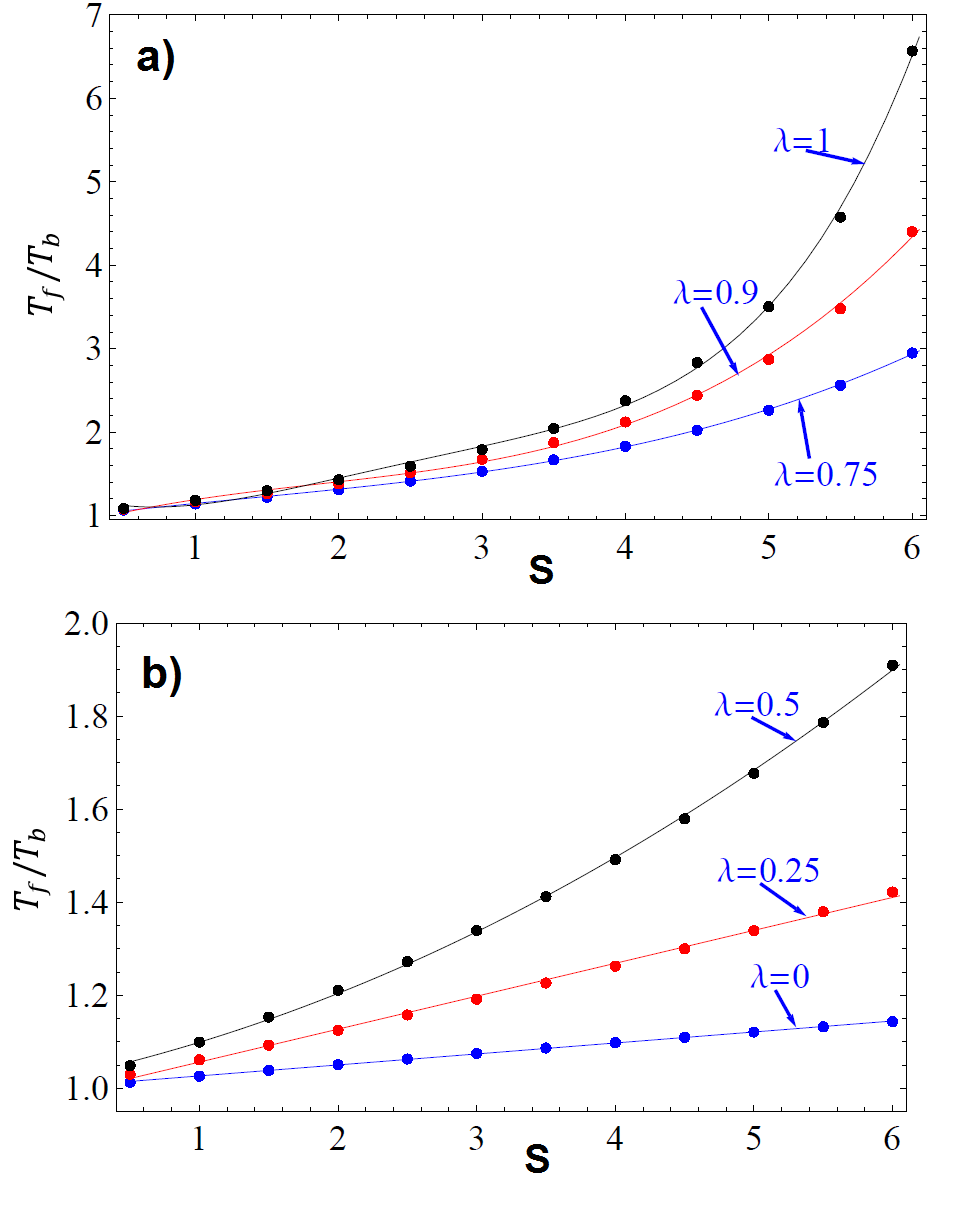}
                \caption{ Dependence of the temperature of the cavity field $T_f$, relative to cold bath temperature $T_b$, 
                at the end of the isothermal expansion on spin-$S$ value for different anisotropy parameters $\lambda$. 
                The parameters are the same as those used in plot in Fig.~\ref{fig:fig1}. The solid curves are the results of a non-linear 
                fit obtained using the function $y=a+bx+cx^2+dx^3+ex^4$ and the fitting parameters are presented in Table $1$.}
   \label{fig:fig2}
\end{figure}

We determine the PCE dynamics by using two different approaches. First, we use a coarse-grained master equation which allows for analytical results for ideal conditions. The second approach is numerical and entails the solution of the microscopic 
master equation, including the atomic decay and cavity loss into account. Initially the 
cavity field is assumed to be in thermal state at $T_b$. 
The model describing the interaction of a tape spin with the  cavity field is given by 
Jaynes-Cummings Hamiltonian~\cite{jaynes_comparison_1963}, 
$H_I=g\left(\sigma_{0+}a+\sigma_{0-}a^{\dagger}\right)$, where $g$ is the coupling constant, 
$\sigma_{0\pm}=\sigma_{0x}\pm i\sigma_{0y}$ are the spin-$1/2$ ladder operators for the tape spin, 
$a$ and $a^{\dagger}$ are the ladder operators of the cavity field with the unperturbed Hamiltonian 
$H_{c}=\Omega a^{\dagger}a$, where $\Omega$ is the cavity field frequency. We consider a resonant
interaction $\omega=\Omega$. Similar to the cycles in Ref.~\cite{scully_extracting_2003,quan_quantum-classical_2006,turkpence_quantum_2016}, we assume 
$\Omega$ and $\omega$ change slightly during the isothermal stages. 

For a large number of pump atoms, the atomic beam can be described as an effective hot bath using a coarse-grained master
equation approach. If the atoms are injected randomly with arrival probability $r$ and an atom interacts with the cavity for a time interval $\tau$ ($r\leq 1/\tau$), then the coarse-grained master equation describing the evolution of the field 
reads~\cite{dillenschneider_energetics_2009}
\begin{eqnarray}
\label{cdm}
\dot\rho_c&=&\frac{\alpha p_e}{2}\!\left(2a^{\dagger}\rho_c a{-}\{\rho_c, a a^{\dagger}\}\right)\nonumber\\
&+&\frac{\alpha(1{-}p_e)}{2}\!\left(2a\rho_c a^{\dagger}{-}\{\rho_c, a^{\dagger}a\}\right),
\end{eqnarray}
where $\alpha=r(g\tau)^2$. Here $\alpha p_e$ and $\alpha(1-p_e)$ are the effective rates for the amplification and damping of the field, respectively. Below maser threshold, which is consistent with the detailed balance condition, the steady state solution of Eq.~(\ref{cdm}) is a thermal state at temperature $T_f=T_q=-\omega/\ln[p_e/(1-p_e)]$. 
Fig.~\ref{fig:fig2} shows how $T_f$ depends on $S=N/2$ for different $\lambda$. A nonlinear behaviour with $S$ emerges and tunable with $\lambda$ as tabulated in Table~\ref{table1}. We observe a monotonic growth at $\lambda=1$ in Fig.~\ref{fig:fig2}(a) and deviation from linear scaling after $\lambda=0.5$. For $\lambda<0.5$, linear scaling is the dominant behavior [cf.~Fig.~\ref{fig:fig2}(b)]. 

The behaviors in Fig.~\ref{fig:fig2}
can be related to the quantum correlations surviving at finite temperatures in Heisenberg XXZ model. 
Though there are results of scaling of zero temperature entanglement in XX model ($\lambda=0$)~\cite{hutton_mediated_2004}, 
calculations of scaling of thermal discord and entanglement for multiple spins
are challenging. Besides, while bipartite discord and concurrence contribute to enhancement of $T_f$ for the case
of two spins~\cite{dillenschneider_energetics_2009}, multipartite correlations mediated by the central spin~\cite{hutton_mediated_2004} 
could contribute as well in the case of large spin bath. {Such central spin-mediated thermal quantum correlations can be useful in light of their robustness against deviations from homogeneous models~\cite{anza_tripartite_2010}. 
The rigorous relation between quantum correlations and the scaling behavior in Fig.~\ref{fig:fig2} as well as its robustness in
inhomogeneous central spin models are beyond our scopes and left as future points to address.}

The study of Fig.~\ref{fig:fig2} shows that a spin-star quantum fuel can lead to more advantageous scaling with the number of
quantum resources than superradiant~\cite{hardal_superradiant_2015}
or multilevel quantum fuels~\cite{turkpence_quantum_2016}. The latter lead to $\sim N^2$ enhancement of $T_f$, while the spin-star allows for $\sim N^3$ or $\sim N^4$, depending on $\lambda$. 
An appealing feature of a PCE is that it can operate with a
single heat bath while the second bath is a quantum resource, albeit with reportedly very low efficiencies
~\cite{scully_extracting_2003,turkpence_quantum_2016}.
The efficiency in the case of the spin-star scheme is several orders of magnitude larger than other PCEs. 
For instance, for $\lambda=1$ and $S=5$, the cavity field temperature is $T_f/T_b=3.28$ corresponding to $\eta\sim 70\%$. 
This estimation however is for ideal conditions, where we ignored the cooling of the cavity by the environment
when it is empty during the initialization of the tape spins. Even if we inject the tape spins into the cavity immediately after they are ejected out of the hohlarum, the thermalization of the spin system would happen in finite time in real systems. In order to estimate how long it would take for thermalisation to occur without loosing benefits of the scaling up the cavity temperature, we resort to a numerical analysis.

We apply the microscopic master equation approach in micromaser dynamics~\cite{briegel_quantum_1993,briegel_one-atom_1994} and assume typical microwave cavity parameters~\cite{blais_cavity_2004} in our numerical simulations. 
Instead of random injection of the atomic beam, we consider regular atomic pump, i.e. atoms entering the cavity at equidistant times and
  an injection rate $r$. In some regimes, pump statistics can influence the variance of photon 
distribution~\cite{bergou_role_1989}. Here we focus on the mean number of cavity photons, which determines the radiation 
pressure and the work output. 
Microscopic master equation approach simulates 
two stages of the micromaser dynamics. In the first one, atoms pass through the cavity one by one: each atom
interacts with the cavity for a short 
time $\tau$. In the second stage, the cavity is empty for duration $\tau_0$ during which the tape spins are initialized and transferred
to the cavity. Therefore, $1/r=\tau+\tau_0$. In order to achieve a steady state we have to ``kick'' the cavity field many
times by repeated atomic injections before the field decays. We thus consider the number $N_{\mathrm{ex}}$ of atoms interacting with the cavity field within the photon lifetime. Clearly, $1/r=1/(N_{\mathrm{ex}}\kappa)$. 
\begin{figure}[!t]
            \includegraphics[width=7.9cm]{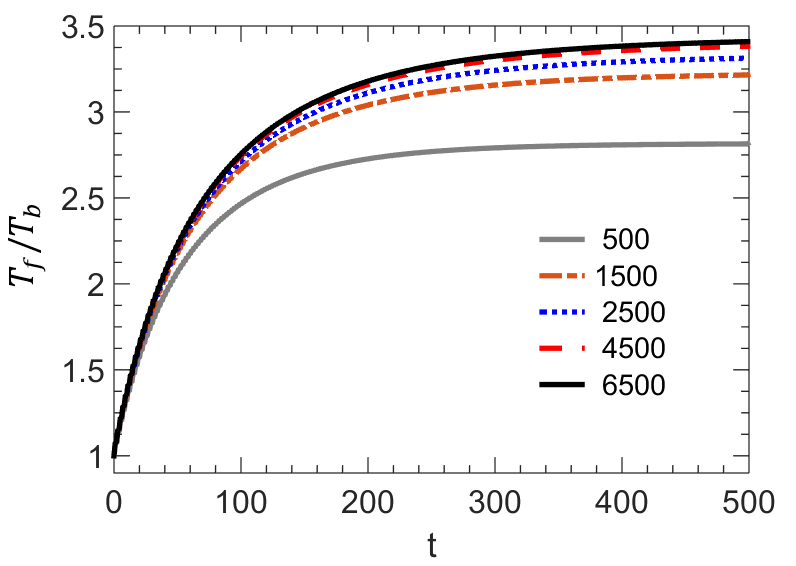}
                \caption{ Time evolution of effective temperature $T_f$ of the cavity field, relative to bath temperature $T_b$, 
                under regular injection of tape spins for different $N_{\mathrm{ex}}=500, 1500, 2500, 4500, 6500$ in increasing order 
                from the lower to upper curves, respectively. Initially, $T_f=T_b$ and the spins are prepared by either of the two 
                initalization schemes in Fig.~\ref{fig:fig1} with $S=N/2=5$ and $\lambda=1.0$ in local thermal states at temperature 
                $T_q/T_b=3.28$. The resonant field frequency is $\Omega/2\pi=50$ GHz, cavity quality factor is 
                $Q=2\times 10^{10}$, transition time through cavity is $ \tau=9.5$ $\mu$s, atomic decay rate is $\gamma/2\pi=33.3 $ Hz 
                and atom-field coupling is $g/\pi=50 $ kHz. Time is dimensionless and scaled with $\Omega$.}   
      \label{fig:fig3}
\end{figure}
\begin{figure}[!t]
            \includegraphics[width=8.8cm]{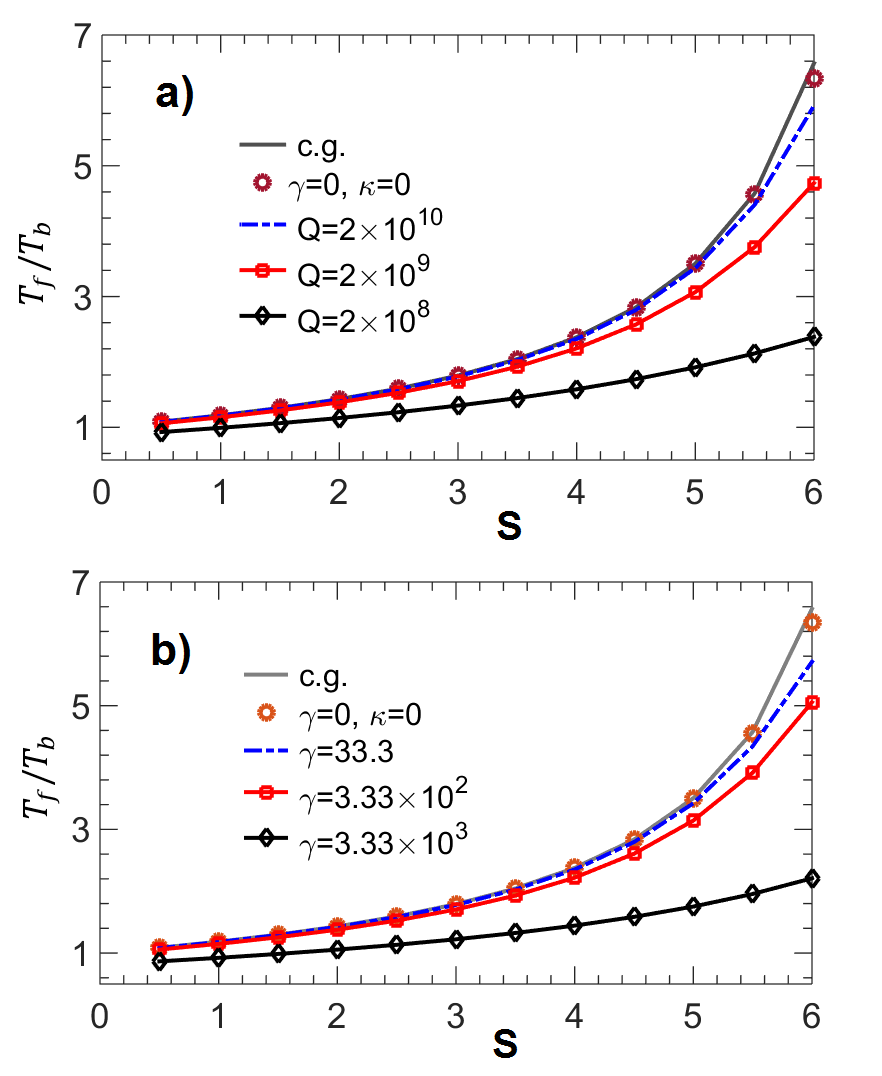}
                \caption{Dependence of the steady state cavity field temperature $T_f$, relative to bath temperature $T_b$, on the number of
                outer spins $N=2S$ for different cavity quality factors $Q$ and atomic decay rates $\gamma$. 
                The curve for the coarse grained master equation case is labeled with  ``c.g.''. Other curves are obtained with the microscopic 
                master equation method. Top and bottom panels present the individual effect of  (a) $\kappa=\Omega/Q$ cavity loss  
                and (b) $\gamma$ atomic decay, respectively. We use $N_{\mathrm{ex}}=6500$ and the other parameters are 
                the same as in Fig.~\ref{fig:fig3}.}
   \label{fig:fig4}
\end{figure}

The initial density matrix of the composite system 
is given by the tensor product  
$\rho=\rho_c(0)\otimes\rho_0$, where $\rho_0$ is the effective atomic thermal state at $T_q$, 
initialized by either of the two schemes, and $\rho_c(0)$ is the 
thermal state of the cavity field at $T_b$. We take the cavity initial state at $T_b$ instead of $T_q$ so as to examine the
effective quantum heating of the cavity by the tape spins. This allows us to see the capabilities
of the spin-star fuel both for quantum thermalization~\cite{liao_single-particle_2010} and for maintaining the cavity at a steady temperature. 
The values of the parameters that would be needed to maintain a steady state of the cavity close to $T_q$ will be used in the initial state of the isothermal-expansion stage.
The initial state is evolved for the short time $\tau$ by the master equation~\cite{briegel_quantum_1993,briegel_one-atom_1994}
\begin{equation}\label{master1}
\dot{\rho}=-i[H,\rho]+\gamma\mathcal{D}[\sigma_{0-}]+\kappa\mathcal{D}[\hat{a}]
\end{equation}
where $H=H_c+(\omega/2)\sigma_{0z}+H_I$ describes the Hamiltonian part of the dynamics 
and $\mathcal{D}[x]=(2x\rho x^{\dagger}-\{xx^{\dagger},\rho\})/2$ is the Liouvillian superoperator  
describing the relaxations to environment reservoir.  The second term is due to the atomic transition occurring without
emitting a photon into the cavity mode. 
After that, the reduced density matrix of the cavity field is evolved by the master equation for the time $\tau_0$~\cite{briegel_quantum_1993,briegel_one-atom_1994}
\begin{equation}\label{master2}
\dot{\rho}_c=-i[H_c,\rho_c]+\kappa\mathcal{D}[{a}_c],
\end{equation}
which characteterizes the damped oscillatory behavior of the empty cavity between successive tape spins. The procedure is repeated by resetting the atomic state and updating the cavity field in the initial state, 
until the steady state of the cavity is reached. The master equations are written under the assumption of $T_b\ll\Omega$.
In our simulations we take $\omega/T_b\sim 6$. For $\omega/2\pi=50$ GHz this gives $T_b\sim 300$ mK. The excitation
number in the cold bath ($\bar n_b\sim 0.002$) can be neglected and the cold bath treated effectively as a zero-temperature
bath. {For simplicity, we use the conditions in Refs.~\cite{scully_extracting_2003,quan_quantum-classical_2006,turkpence_quantum_2016},
where  the variation of $\Omega$ in isothermal stages is much
smaller than $\Omega$ so that the parameters $\omega,\Omega,\gamma,\kappa$ can be considered constant in successive iterations of Eqs.~(\ref{master1}) and~(\ref{master2}).}
In addition, we neglect the small atomic  dephasing term ($\gamma_\phi/2\pi\sim 3.3$ Hz).

We calculate the steady state value of mean photon number in the cavity 
$\bar{n}_c=\mathrm{Tr}(\rho_c{a}^{\dagger}{a})$ which is used to determine 
$T_f=\Omega/\ln(1+1/\bar{n}_c)$. 
The field temperature $T_f$ is effectively defined in the presence of 
cavity losses~\cite{pielawa_engineering_2010,liao_single-particle_2010}. 
It gets closer to a genuine temperature when
$\tau\ll 1/g$~\cite{liao_single-particle_2010}. We have $g/\pi=50$ kHz which leads to $\tau\ll 40\,\mu$s. We fix $\tau=9.5\,\mu$s
and vary $\tau_0$ through $N_{\mathrm{ex}}$. 
In Fig.~\ref{fig:fig3} we present the dynamics of $T_f$ depending on different $N_{\mathrm{ex}}$ values. 
Initially the cavity is in thermal equilibrium with the cold bath $T_f/T_b=1$. After injection of many spins at 
$T_q>T_b$, $T_f$ increases and reaches a steady state value, which is smaller than $T_q$ due to losses.  
The coarse-grained master
equation predicts a thermal state of the cavity at $T_f=T_q$. By 
increasing $N_{\mathrm{ex}}$, the cavity interacts with more spins in a photon lifetime. 
Accordingly, the loss can be compansated. Around $N_{\mathrm{ex}}\sim 6500$ numerical
behavior of $T_f$ conforms to the analytical predictions. At this case,
time between succesive spins becomes $1/r\sim 10\,\mu$s for $\kappa\sim 5\pi$ Hz. 
The cavity loss $\kappa=\omega/Q$ is evaluated for
a high finesse cavity with quality factor $Q=2\times 10^{10}$ at a microwave cavity mode frequency $\omega/2\pi=50$ GHz. 
As we fixed the transition time at $\tau_0=9.5\,\mu$s, this leaves about $\tau_0\sim 500$ ms for thermalization
of the spin-star network and to take away the central spin to the cavity. If this time is found to be
too short in a particular implementation, one can use an ensemble of spin-star networks, instead of a single 
one, and collect the central spins on demand to generate a convenient injection time distribution. 

The harmful effect of increasing the empty cavity time $\tau_0$, or equivalently decreasing $N_{\mathrm{ex}}$ can be studied by considering larger relaxation parameters. By fixing $\gamma$ and increasing $\kappa$,
the advantages of scaling in $T_f$ with $S=N/2$ diminish, as shown in Fig.~\ref{fig:fig4} (a).
Similar conclusion can be deduced for increasing $\gamma$ at fixed $\kappa$ in Fig.~\ref{fig:fig4} (b); besides, the 
sensitivities of the system to 
either atomic or cavity losses are similar, as revealed by comparing the top and bottom panels in Fig.~\ref{fig:fig4}. Robustness of nonlinearity under strong damping
can be exploited to tolerate longer thermalization and transfer times. We conclude that spin-star network-powered
PCE can attain high efficiencies, even in the presence of decoherence. 
\section{Conclusions}
We proposed the use of an $N$-element spin-star network as a quantum fuel to power up a PCE. 
Such a choice offers significant advantages embodied by substantial quantum coherences generated and preserved naturally, in the form of thermal entanglement. When the network is in thermal equilibrium with a heat bath, the outer spins are in a state of non-equilibrium, and the central spin is in an effective thermal state with a temperature exhibiting superlinear dependence on $N$ and higher than the one of the bath. Owing to the thermal nature of the central spin, large quantum coherences can be exploited to power up a PCE  
operating in contact with a single heat bath, even in the presence of atomic and cavity damping. We have found an efficiency of $\sim 70\%$, which is several orders of magnitude larger than typical PCEs. In addition, we have discussed an equivalent scheme, where the outer spins are replaced by a single macroscopic spin. Various physical platforms, from quantum dots to molecular nanomagnets, superconducting spins, NMR and circuit-QED, can be considered for a proof-of-principle realization.
\acknowledgments
\"{O}.~E.~M.~gratefully acknowledges support by 
Ko\c{c} University (KU) Visiting
Scholar Program by KU Office of VPAA (Vice President
of Academic Affairs) and  hospitality at Queen's University Belfast. M. P. is supported by the
EU FP7 grant TherMiQ (Grant Agreement 618074), and the J. Schwinger Foundation (grant number JSF-14-7-0000).
The authors acknowledge support from the University Research Agreement between Ko\c{c} University and Lockheed
Martin Corporation and the Royal Society-Newton Mobility Grant NI160057.

\end{document}